\documentclass[aip,reprint,jap]{revtex4-1}

\usepackage{verbatim}
\usepackage{amsmath}
\usepackage[pdftex]{graphicx}
\usepackage[thinspace]{SIunits}
\usepackage[english]{babel}
\usepackage{blindtext}

\begin{document}

\title{A carrier relaxation bottleneck probed in single InGaAs quantum dots using integrated superconducting single photon detectors}

\author{G.~Reithmaier}\email{guenther.reithmaier@wsi.tum.de}\affiliation{Walter Schottky Institut, Technische Universit\"at M\"unchen, Germany}
\author{F.~Flassig}\affiliation{Walter Schottky Institut, Technische Universit\"at M\"unchen, Germany}
\author{P.~Hasch}\affiliation{Walter Schottky Institut, Technische Universit\"at M\"unchen, Germany}
\author{S.~Lichtmannecker}\affiliation{Walter Schottky Institut, Technische Universit\"at M\"unchen, Germany}
\author{K.~M\"uller}\affiliation{Walter Schottky Institut, Technische Universit\"at M\"unchen, Germany}\affiliation{E L Ginzton Laboratory, Stanford University, Stanford, CA 94305, USA}
\author{J.~Vu\v{c}kovi\'{c}}\affiliation{E L Ginzton Laboratory, Stanford University, Stanford, CA 94305, USA}\affiliation{TUM Institute of Advanced Study, Lichtenbergstraße 2a,85748 Garching, Germany}
\author{R.~Gross}\affiliation{Walther Mei{\ss}ner Institut, Bayerische Akademie der Wissenschaften, Garching, Germany}\affiliation{Nanosystems Initiative Munich (NIM), Schellingstraße 4, 80799 München, Germany}
\author{M.~Kaniber}\affiliation{Walter Schottky Institut, Technische Universit\"at M\"unchen, Germany}
\author{J.~J.~Finley}\email{finley@wsi.tum.de}\affiliation{Walter Schottky Institut, Technische Universit\"at M\"unchen, Germany}\affiliation{Nanosystems Initiative Munich (NIM), Schellingstraße 4, 80799 München, Germany}

\date{\today}

\begin{abstract}

Using integrated superconducting single photon detectors we probe ultra-slow exciton capture and relaxation dynamics in single self-assembled InGaAs quantum dots embedded in a GaAs ridge waveguide. Time-resolved luminescence measurements performed with on- and off-chip detection reveal a continuous decrease in the carrier relaxation time from $1.22 \pm 0.07$ ns to $0.10 \pm 0.07$ ns upon increasing the number of non-resonantly injected carriers. By comparing off-chip time-resolved spectroscopy with spectrally integrated on-chip measurements we identify the observed dynamics in the rise time ($\tau_r$) as arising from a relaxation bottleneck at low excitation levels. From the comparison with the temporal dynamics of the single exciton transition with the on-chip emission signal, we conclude that the relaxation bottleneck is circumvented by the presence of charge carriers occupying states in the bulk material and the two-dimensional wetting layer continuum. A characteristic $\tau_r \propto P^{-2/3}$ power law dependence is observed suggesting Auger-type scattering between carriers trapped in the quantum dot and the two-dimensional wetting layer continuum which circumvents the phonon relaxation bottleneck.

\end{abstract}

\maketitle

Semiconductor based photonic information technology is rapidly being pushed to the quantum limit where single photon states can be generated and manipulated in nanoscale optical circuits\cite{Matthews09}. Over recent years quantum dots (QDs) embedded in such semiconductor systems have been shown to be excellent sources of quantum light\cite{Flagg09,Gao12,He13} and have shown their suitability for use as a gain medium in QD lasers\cite{Ellis11,Ledentsov98,Ellis07}. However, for short response times and fast operation of such devices, injected charge carriers must relax rapidly from continuum wetting layer electronic states into the lasing state. For a fully discrete electronic structure, efficient relaxation is expected to be hindered by phonon bottleneck phenomena\cite{Urayama01}, caused by the large energetic spacing of QD energy levels that inhibits single-phonon mediated scattering processes\cite{Benisty91}. To directly observe such relaxation bottleneck effects, superconducting single photon detectors (SSPDs) are suitable due to their near unity quantum efficiency and picosecond timing resolution\cite{Goltsman05,Najafi12}. Building up on recent progress in this field\cite{Pernice12,Sprengers11,Sahin13}, we developed highly efficient\cite{Marsili08,Kerman07} NbN-SSPDs on GaAs\cite{Reithmaier13} and demonstrated the monolithic integration of InGaAs QDs as single photon emitters together with waveguides and detectors on a single chip\cite{Reithmaier13SciRep}.

\begin{figure}[h!]
\includegraphics{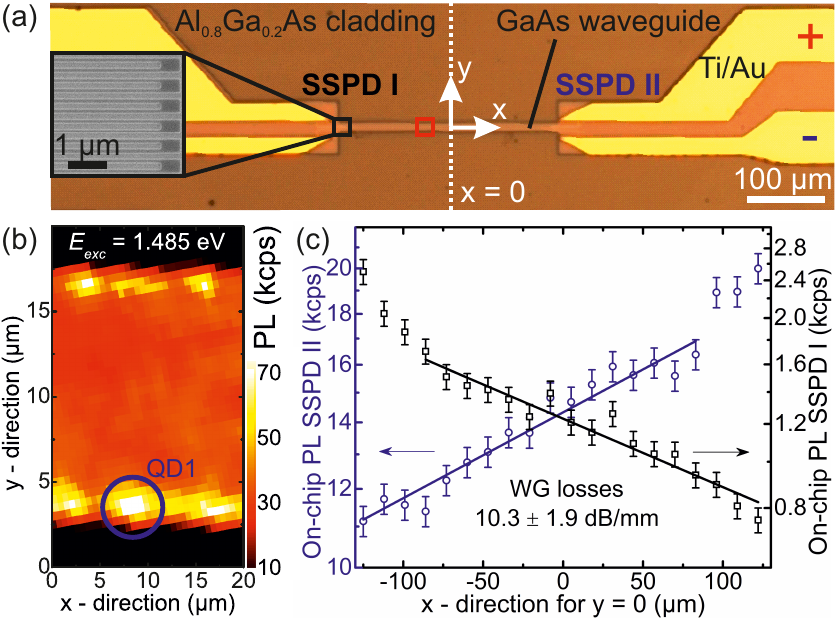}
\caption{(a) Sample design showing the GaAs waveguide grown on a Al$_{0.8}$Ga$_{0.2}$As cladding with the two SSPDs and Ti/Au contact pads. Inset: SEM image of lithographically defined NbN nanowires. (b) On-chip PL scan across the red-marked region of the waveguide. (c) On-chip PL of SSPD I (black squares) and SSPD II (blue circles) as a function of position along the waveguide, shown on a semi-logarithmic scale.}
\label{fig1}
\end{figure}

In this letter, we compare photoluminescence (PL) dynamics recorded from a single dot with confocal off-chip detectors with on-chip PL using integrated SSPDs that provide temporal resolution better than $70$ ps. By probing the carrier capture and energy relaxation dynamics of photogenerated charge carriers, we demonstrate that the rise time of the luminescence signal ($\tau_r$) is strongly dependent on the excitation power level. The ultrafast response and high detection efficiency of the integrated SSPDs enables us to probe the PL dynamics at low excitation levels. In such a scenario carrier capture and energy relaxation is inefficient due to the discrete electronic structure of the dots and the absence of Coulomb mediated scattering involving carriers occupying energetically higher discrete states\cite{Urayama01}. In particular, an excitation power dependent change in inter-sublevel relaxation times is observed for a single, spatially isolated QD. Auger-like scattering involving carriers occupying excited QD energy states and proximal continuum states is shown to be the dominant mechanism for the vanishing of this relaxation bottleneck\cite{Urayama01} at elevated excitation powers. The simultaneous off-chip PL measurement of the spectrally filtered neutral exciton ($X_0$) transition exhibits a plateau in the temporal response of the system, consistent with the change in carrier relaxation as well as the saturation behaviour of the neutral exciton emission. Finally, measurements of the carrier capture and relaxation time as a function of the single exciton population probability reveal that the observed effect is universal for all excitation energies studied.

The sample studied consists of a single layer of epitaxially grown, self-assembled InGaAs QDs with an areal density of $2-3$ $\mu$m$^{-2}$ embedded in a $15$ $\mu$m wide GaAs-Al$_{0.8}$Ga$_{0.2}$As multimodal ridge waveguide. As presented in detail elsewhere\cite{Reithmaier13SciRep}, the QDs have a typical lateral (vertical) size of $25 \pm 5$ nm ($5 \pm 1$ nm) and are embedded at the midpoint of the $250$ nm thick GaAs waveguide core and photolithography and wet chemical etching are applied to define linear waveguide structures investigated. Light emitted by the QDs is guided along the $275$ $\mu$m long waveguide and evanescently coupled\cite{Sahin13, Reithmaier13SciRep,Pernice12,Sprengers11} into superconducting NbN detectors at the remote ends. Each end of the waveguide is equipped with a separate detector, labeled SSPD I and SSPD II in figure \ref{fig1}a. Measurements of the optical function of the NbN film combined with finite difference time domain (FDTD) simulations indicates that $\sim 97.8\%$ of the light is absorbed by the array of $80 \pm 10$ nm wide NbN nanowires\cite{Reithmaier13SciRep} over the total light - detector interaction length of 23 $\mu$m. When operated at cryogenic temperatures ($T = 4.2$ K) the NbN SSPDs were found to be single photon sensitive in the near infrared with an average dark count rate $\leq 1$ cps. When scanning a laser across the midpoint of the waveguide, with an energy resonant with the wetting layer continuum ($E_{exc} = 1.485$ eV) and a sufficiently high power density to create multiple excitons per dot ($P = 172 \pm 7$ W/cm$^2$), we observe $50 \pm 15$ kcps of on-chip PL with SSPD II at an average background count rate of only $10 \pm 1$ kcps arising from scattered laser light, as presented in figure \ref{fig1}b. The resulting PL map shows well defined bright spots corresponding to single dots that can be excited most effectively when they are located close to the edge of the waveguide. This is attributed to an enhanced incoupling of light via the sloping waveguide edges at the air-GaAs interface, arising from the isotropic wet-chemical etching process. Recording the on-chip PL for similar excitation conditions as a function of the laser spot position along the +x direction on the waveguide, we observe an exponential decrease/increase of the signal for SSPD I/II in the semi-logarithmic plot in figure \ref{fig1}c, from which we determine the waveguide losses to be $10.3 \pm 1.9$ dB/mm. The difference between the total count rates recorded for both detectors stems from fabrication imperfections, leading to a $\sim 10 \times$ higher detection efficiency for SSPD II compared to SSPD I.

\begin{figure}[h!]
\includegraphics{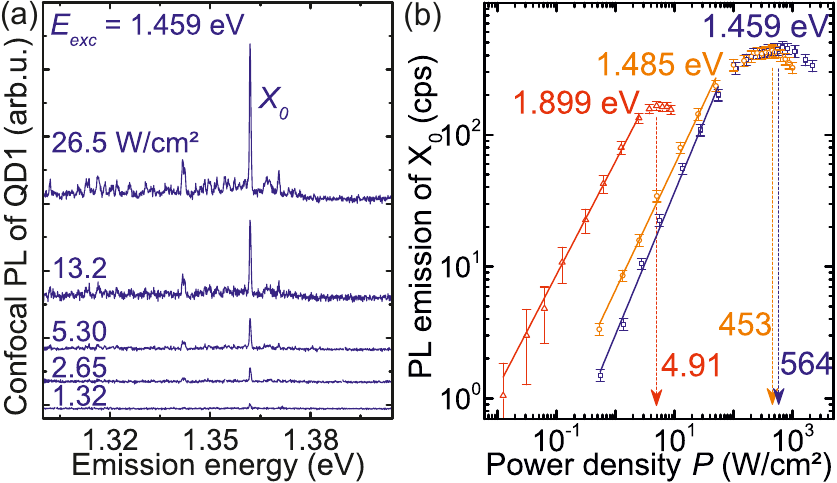}
\caption{(a) Low power PL spectra of QD1 for an excitation energy of $E_{exc} = 1.459$ eV. (b) Power dependent PL of $X_0$ for three different excitation energies.}
\label{fig2}
\end{figure}


For the following measurements we focused the excitation laser onto one individual dot, positioned at a distance $x_0=150$ $\mu$m ($x_0'=200$ $\mu$m) from SSPD I (SSPD II) and labelled QD1 in figure \ref{fig1}b. To study the emission spectrum of QD1, we performed power dependent PL spectroscopy recorded in a confocal geometry.  Upon varying the excitation power density from 1.32 W/cm$^2$ up to 26.5 W/cm$^2$ a single, dominant emission line with a resolution limited linewidth of $0.53 \pm 0.02$ meV is observed in the emission spectra shown in figure \ref{fig2}a, attributed to a single exciton transition $X_0$ by its clear linear power dependence. The weak emission lines in the vicinity stem from the emission of either nearby quantum dots or charged excitonic transitions\cite{Finley01} of QD1. The evolution of the peak amplitude of the $X_0$ emission for three different excitation energies is presented in figure \ref{fig2}b. In this double logarithmic plot, a linear increase of the quantum dot emission with an exponent of $0.95 \pm 0.02$ is observed for all excitation energies, supporting our identification of the single excitonic character of the studied emission line\cite{Finley01}. Depending on the excitation energy, the emission of $X_0$ saturates at $4.91$ W/cm$^2$, $453$ W/cm$^2$ and $564$ W/cm$^2$ for $E_{exc} = 1.899$ eV, $1.531$ eV and $1.459$ eV, respectively, as marked in figure \ref{fig2}b. Here, the saturation power density is extracted at the maximum of the PL emission. The lower maximum PL intensity for excitation at $1.899$ eV is attributed to an increased charged exciton population of the dot, arising from the higher absorption stength and the increased carrier diffusion length at this excitation energy. The trend of an increased saturation power density with decreasing excitation energy reflects the stronger absorption at higher photon energies and the accompanying larger density of photo-generated charge carriers for a specific excitation level\cite{Vasanelli02}. Similar effects are found to result in faster carrier capture dynamics for higher laser excitation energies in our time resolved studies presented below.

\begin{figure*}
\includegraphics{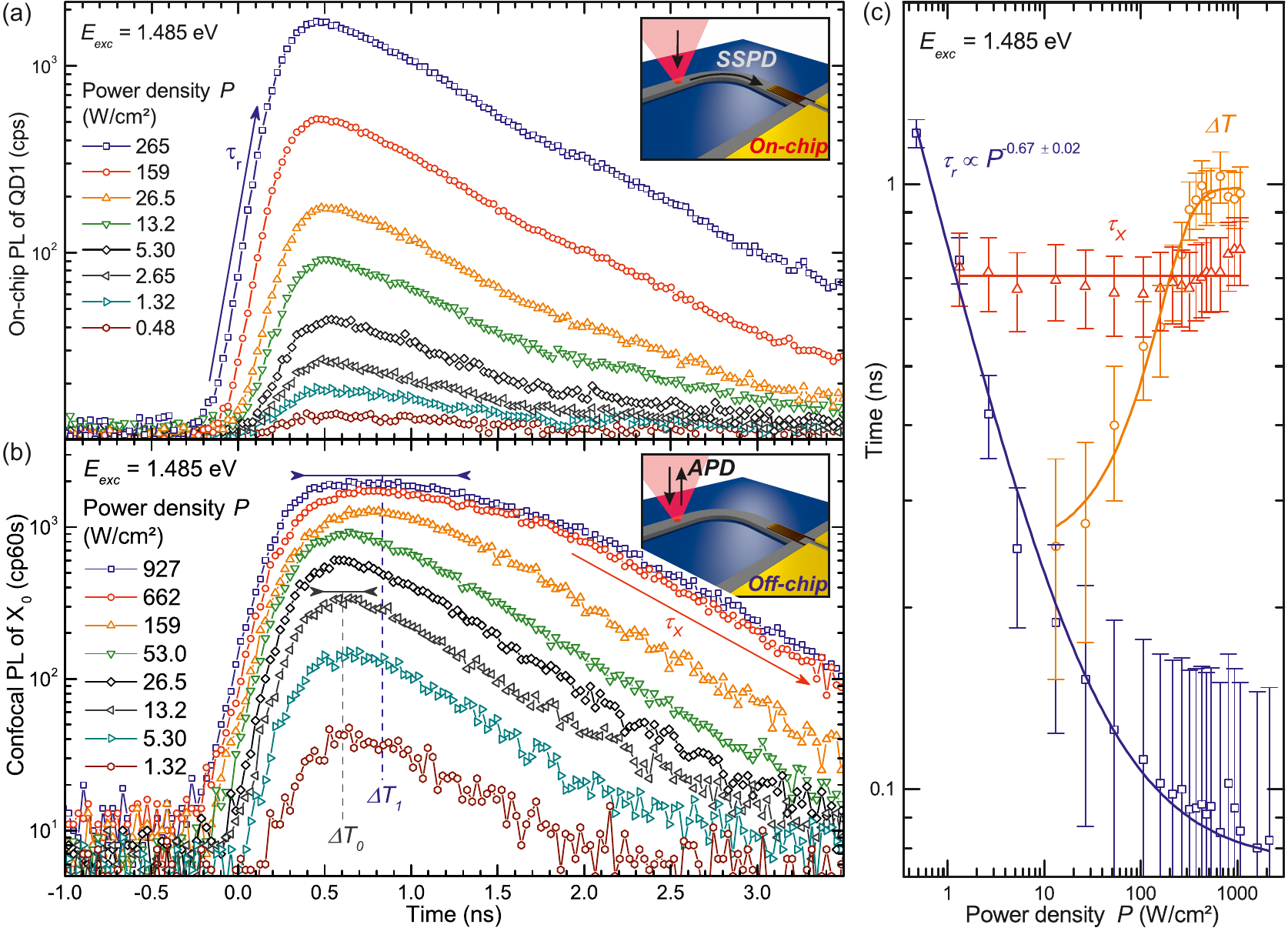}
\caption{(a) Time-resolved on-chip PL with the laser excitation positioned on QD1 presented for excitation power densities $P$ ranging from 0.48 to 265 W/cm$^2$. Fitting the rising edge of the decay transient, we extract a charge carrier relaxation time of $\tau_{r}=1.22 \pm 0.07$ ns for the lowest power density and $0.10 \pm 0.07$ ns for the highest one, respectively. (b) Spectrally filtered time-resolved confocal PL studies of $X_0$, recorded for power densities increasing from $P = 1.32$ to $927$ W/cm$^2$. The exciton lifetime is determined to be $\tau_X = 0.69 \pm 0.1$ ns. A plateau region $\Delta T_0 = 0.25 \pm 0.10$ ns forms at $P = 13.2$ W/cm$^2$ and increases to $\Delta T_1 = 0.95 \pm 0.10$ ns for the highest excitation power density. (c) Carrier relaxation time $\tau_r$ (blue squares), plateau length $\Delta T$ (orange circles) and exciton lifetime $\tau_X$ (red triangles) as a function of the excitation power density $P$.}
\label{fig3}
\end{figure*}


Employing on-chip PL detection\cite{Reithmaier13SciRep} enables us to study the charge carrier capture and relaxation dynamics of QD1 as a function of the excitation power density, schematically indicated by the icon in the top right of figure \ref{fig3}a. Typical results from such experiments are presented in a semi-logarithmic plot in figure \ref{fig3}a. Here, a tunable TiSa laser with a pulse width $< 1$ ps was focussed onto QD1, highlighted in figure \ref{fig1}b, and time resolved decay transients were recorded using SSPD I. Due to the fast intrinsic timing resolution of the SSPDs ($<70$ ps\footnote{The overall temporal resolution of the on-chip experiment is limited to $< 70$ ps, reached by deconvolution of the signal being recorded using a Picoquant Timeharp 200 that exhibits a timing uncertainty of $< 150$ ps.}), this measurement technique allows us to directly probe carrier capture, relaxation and recombination dynamics of QD1. The data presented in figure \ref{fig3}a is shown for different excitation levels, excited into the WL continuum at $E_\text{exc} = 1.485$ eV. For the lowest power density used of $P = 0.48$ W/cm$^2$ (dark red circles in figure \ref{fig3}a) we observe an exponential increase followed by an almost perfect mono-exponential decay at longer timescales \cite{Schwoob05,Rao07}. For higher excitation powers, exceeding $26.5$ W/cm$^2$ in figure \ref{fig3}a (orange triangles), the decay becomes multi-exponential, as light emitted by the faster decays\cite{Santori02,Raymond96} of multi-excitonic transitions is detected by the SSPD as well as that from $X_0$, since the SSPD does not spectrally resolve the emission. More remarkably, while increasing $P$ up to $265$ W/cm$^2$, the rising edge gradually becomes steeper and the relaxation time $\tau_r$ reduces from its initial value of $1.22 \pm 0.07$ ns to $0.10 \pm 0.07$ ns, shown by the dark blue squares in figure \ref{fig3}a. This behaviour suggests a power-dependent change of the carrier capture and inter-sublevel relaxation times which are typically governed by Auger-like processes or LO-phonon emission in QDs\cite{Raymond96} and will be discussed in relation to figures \ref{fig3}c and \ref{fig4} below. The ultraslow carrier relaxation time for $P = 0.48$ W/cm$^2$ indicates the presence of an intrinsic phonon bottleneck\cite{Urayama01} that, for stronger excitation levels, is masked by much faster carrier capture and relaxation via Auger-like processes with other charge carriers in the near vicinity of the quantum dot\cite{Mghaieth99}. In similar off-chip detected time-resolved PL experiments, $\tau_r$ typically is $\leq 100$ ps\cite{Raymond96} and does not show any indications of a bottleneck. In our case, the highly selective excitation of QD1, the ultra-low excitation power densities $\leq 1$ W/cm$^2$ in combination with the excellent timing resolution of the SSPD, enables us to observe and systematically investigate the phonon-bottleneck via the variation of the luminescence timescale.


In order to show that faster inter-sublevel relaxation is indeed mediated by scattering with charge carriers occupying higher states in QD1, spectrally and time-resolved PL measurements were performed in a confocal geometry by spectral filtering of the $X_0$ emission. The timing resolution in this experiment is limited by the avalanche photodiode (APD) to $\sim 100$ ps. The results obtained from the off-chip experiment (inset) are presented in figure \ref{fig3}b. Here, the spectrally resolved PL time-transients are shown on a semi-logarithmic plot for excitation power densities increasing from $1.32$ W/cm$^2$ up to $927$ W/cm$^2$ and a fixed excitation energy in the WL continuum at $1.485$ eV. Again, we observe an exponential increase followed by a mono-exponential decay. In contrast to the on-chip measurements presented in figure \ref{fig3}a, the exclusive detection of the $X_0$ transition gives rise to a power-independent mono-exponential decay with $\tau_X = 0.69 \pm 0.10$ ns (indicated by the red arrow on figure \ref{fig3}b) even up to power densities as high as $P = 927$ W/cm$^2$. However, for elevated power densities the decay starts at a later point in time, such as $0.8 \pm 0.1 \ (1.4 \pm 0.1)$ ns in case of $13.2 \ (927)$ W/cm$^2$ shown by the grey triangles (blue squares) in figure \ref{fig3}b. This delay of $\Delta T_\text{0 (1)} = 0.25 \pm 0.1 \ (0.95 \pm 0.1)$ ns, indicated by the grey/blue double arrow on figure \ref{fig3}b, is related to the cascaded recombination of multi-excitonic states prior to the $X_0$ recombination. Strikingly, the appearance of the plateau at $P \sim 13.2$ W/cm$^2$ coincides with the saturation of $\tau_r$, as shown by the curves for $P \geq 13.2$ W/cm$^2$ in figure \ref{fig3}a.


For a quantitative analysis of this process the plateau length $\Delta T$, the $X_0$ lifetime $\tau_X$ and the carrier relaxation time $\tau_r$ are presented in figure \ref{fig3}c as a function of $P$. For increasing power densities a clear decrease of $\tau_r$ from $1.22 \pm 0.07$ ns down to $0.10 \pm 0.07$ ns is observed, being in good qualitative agreement with recent findings obtained in differential transmission\cite{Steinhoff13}. This trend is followed by a saturation of $\tau_r$ just above the temporal resolution limit of the SSPD for power densities larger than $\sim 10$ W/cm$^2$, while the plateau starts to form with $\Delta T$ gradually increasing from $0.25 \pm 0.1$ ns up to $0.95 \pm 0.1$ ns. The exciton lifetime $\tau_X = 0.69 \pm 0.1$ ns remains constant within the error over the whole range of power densities studied here. The decrease in relaxation time and the simultaneous formation of the Pauli-plateau strongly suggests an interplay between faster carrier relaxation and occupation of multiexcitonic states\cite{Raymond96}. For $\tau_r$ it is found that the data points follow a $P^{-0.67\pm0.02}$ dependency strongly indicating that Auger processes mediate intersublevel relaxation\cite{Mghaieth99}. This observation constitutes strong evidence that the occupation of the dot with multiple carriers mediates faster carrier capture and energy relaxation. As presented in figure \ref{fig4}, this power law and saturation behaviour is found to be universal within the experimental error for all excitation energies studied.

\begin{figure}[h!]
\includegraphics{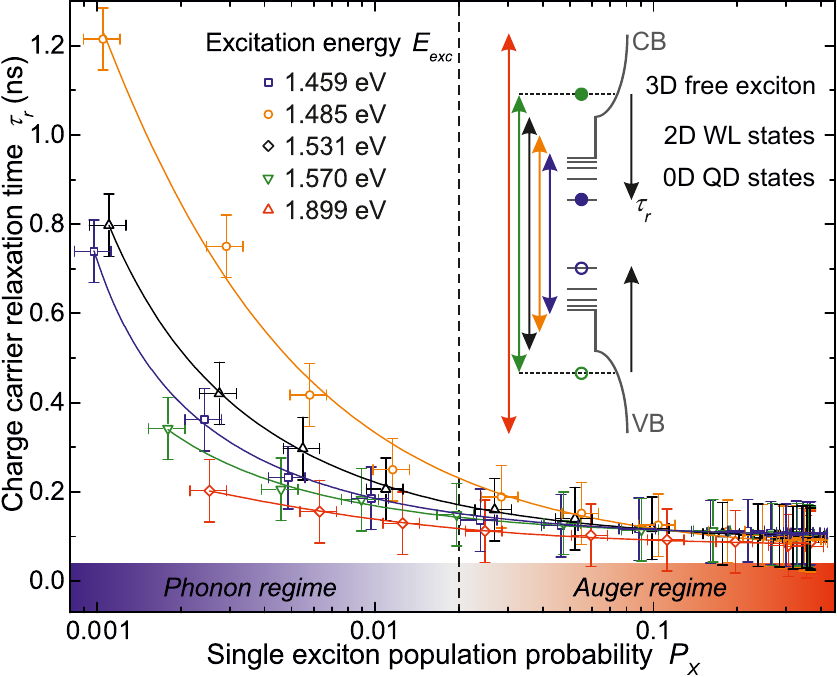}
\caption{Relaxation time $\tau_r$ as a function of the single exciton population probability $P_X$, as determined using Poisson statistics. The lines represent power-law fits with an exponent of $-0.67\pm0.02$. The data is shown for different excitation energies, indicated by the colored arrows in the inset. Here, the density of states of a single QD, grown on a wetting layer and surrounded by GaAs is depicted schematically.}
\label{fig4}
\end{figure}


To gain a deeper insight into the origin of the charge carriers responsible for masking the capture and relaxation bottleneck at elevated power densities, excitation energy dependent PL studies were performed. Here, the excitation power dependent rise time of the luminescence signal $\tau_r$ was extracted for a range of excitation laser energies; starting from the 2D wetting layer continuum at $E_{exc} = 1.459$ eV and increasing up to  $1.899$ eV, far above the GaAs bandgap. To correct for the energy dependent absorption stength, the power density $P$ used for this experiment was converted into an energy independent population probability $P_X$ for a single exciton occupying the QD. Hereby, Poisson statistics \cite{Reithmaier13SciRep,Bacher99} was used with $P_X = \alpha \cdot \exp(-\alpha)$, where $\alpha=P/P_0$ and $P_0$ being the saturation power density, as extracted from the measurements presented in figure \ref{fig2}b. The results of these studies are presented in figure \ref{fig4}. Here, the different excitation energies are indicated by the colored double-arrows on figure \ref{fig4} (inset), that schematically depicts the density of states for valence (VB) and conduction band (CB) of a single dot. $\tau_r$ is indicated by the black arrow showing the relaxation of an electron-hole pair created at $1.570$ eV. Similar to the data recorded for a fixed $E_{exc} = 1.485$ eV (figure \ref{fig3}c and orange circles on figure \ref{fig4}), $\tau_r$ is $1.22 \pm 0.07$ ns for a very low exciton population probability of $P_X = 0.0011 \pm 0.0002$ and then quickly drops towards the resolution limit as $P_X$ increases and is fully saturated at $P_X \sim 0.1$. A similar trend is observed for all laser energies studied, further supporting the identification of two distinct regimes, marked in figure \ref{fig4}. In the \textit{phonon-regime} carrier relaxation is mediated by the emission of acoustic phonons and is inefficient due to the weak matrix element for exciton-phonon coupling\cite{Ikeda04,Heitz01}. In this regime the temperature dependence of charge carrier relaxation proved the involvement of phonons using a Bose-Einstein distribution\cite{Ikeda04}. In our case, the temperature dependence could not be analysed, as the dark counts of the integrated SSPD would exceed the PL signal for temperatures larger than $\sim 5$ K\cite{Kitaygorsky07}. For exciton population probabilities $\geq 0.02$, the \textit{Auger-regime}, carriers rapidly lose excess energy by few-carrier interactions where an e-h pair transfers excess energy non-radiatively to a nearby charge carrier\cite{Mghaieth99}.
  

In summary, by comparing on- and off-chip time-resolved luminescence, we directly probed ultra-slow charge carrier capture and relaxation in a single QD. A continuous decrease in the exciton capture and relaxation times by one order of magnitude was observed upon increasing the number of injected charge carriers. At the same time, a plateau in the QD luminescence with a duration of $0.95 \pm 0.10$ ns is formed, as detected in time-resolved off-chip measurements. This was attributed to the masking of the relaxation bottleneck by the presence of charges in the bulk material and the 2D wetting layer continuum. The universal $P^{-0.67 \pm 0.02}$ power law dependence of the relaxation time with the number of injected charge carriers for all excitation energies studied was attributed to Auger-type scattering, being the dominant effect inhibiting phonon bottleneck phenomena in such systems.


We gratefully acknowledge D. Sahin, A. Fiore (TU Eindhoven), K. Berggren, F. Najafi (MIT), R. Hadfield (Heriot-Watt) and W. Pernice (KIT) for useful discussions and the BMBF via QuaHL-Rep Project number 16BQ1036 and the Bavaria California Technology Center (BaCaTeC) for financial support.


\begin{thebibliography}{31}%
\makeatletter
\providecommand \@ifxundefined [1]{%
 \@ifx{#1\undefined}
}%
\providecommand \@ifnum [1]{%
 \ifnum #1\expandafter \@firstoftwo
 \else \expandafter \@secondoftwo
 \fi
}%
\providecommand \@ifx [1]{%
 \ifx #1\expandafter \@firstoftwo
 \else \expandafter \@secondoftwo
 \fi
}%
\providecommand \natexlab [1]{#1}%
\providecommand \enquote  [1]{``#1''}%
\providecommand \bibnamefont  [1]{#1}%
\providecommand \bibfnamefont [1]{#1}%
\providecommand \citenamefont [1]{#1}%
\providecommand \href@noop [0]{\@secondoftwo}%
\providecommand \href [0]{\begingroup \@sanitize@url \@href}%
\providecommand \@href[1]{\@@startlink{#1}\@@href}%
\providecommand \@@href[1]{\endgroup#1\@@endlink}%
\providecommand \@sanitize@url [0]{\catcode `\\12\catcode `\$12\catcode
  `\&12\catcode `\#12\catcode `\^12\catcode `\_12\catcode `\%12\relax}%
\providecommand \@@startlink[1]{}%
\providecommand \@@endlink[0]{}%
\providecommand \url  [0]{\begingroup\@sanitize@url \@url }%
\providecommand \@url [1]{\endgroup\@href {#1}{\urlprefix }}%
\providecommand \urlprefix  [0]{URL }%
\providecommand \Eprint [0]{\href }%
\providecommand \doibase [0]{http://dx.doi.org/}%
\providecommand \selectlanguage [0]{\@gobble}%
\providecommand \bibinfo  [0]{\@secondoftwo}%
\providecommand \bibfield  [0]{\@secondoftwo}%
\providecommand \translation [1]{[#1]}%
\providecommand \BibitemOpen [0]{}%
\providecommand \bibitemStop [0]{}%
\providecommand \bibitemNoStop [0]{.\EOS\space}%
\providecommand \EOS [0]{\spacefactor3000\relax}%
\providecommand \BibitemShut  [1]{\csname bibitem#1\endcsname}%
\let\auto@bib@innerbib\@empty
\bibitem [{\citenamefont {Matthews}\ \emph {et~al.}(2009)\citenamefont
  {Matthews}, \citenamefont {Politi}, \citenamefont {Stefanov},\ and\
  \citenamefont {O'Brien}}]{Matthews09}%
  \BibitemOpen
  \bibfield  {author} {\bibinfo {author} {\bibfnamefont {J.}~\bibnamefont
  {Matthews}}, \bibinfo {author} {\bibfnamefont {A.}~\bibnamefont {Politi}},
  \bibinfo {author} {\bibfnamefont {A.}~\bibnamefont {Stefanov}}, \ and\
  \bibinfo {author} {\bibfnamefont {J.~L.}\ \bibnamefont {O'Brien}},\
  }\href@noop {} {\bibfield  {journal} {\bibinfo  {journal} {Nat. Phot.}\
  }\textbf {\bibinfo {volume} {3}},\ \bibinfo {pages} {346} (\bibinfo {year}
  {2009})}\BibitemShut {NoStop}%
\bibitem [{\citenamefont {Flagg}\ \emph {et~al.}(2009)\citenamefont {Flagg},
  \citenamefont {Muller}, \citenamefont {Robertson}, \citenamefont {Founta},
  \citenamefont {Deppe}, \citenamefont {Xiao}, \citenamefont {Ma},
  \citenamefont {Salamo},\ and\ \citenamefont {Shih}}]{Flagg09}%
  \BibitemOpen
  \bibfield  {author} {\bibinfo {author} {\bibfnamefont {E.}~\bibnamefont
  {Flagg}}, \bibinfo {author} {\bibfnamefont {A.}~\bibnamefont {Muller}},
  \bibinfo {author} {\bibfnamefont {J.}~\bibnamefont {Robertson}}, \bibinfo
  {author} {\bibfnamefont {S.}~\bibnamefont {Founta}}, \bibinfo {author}
  {\bibfnamefont {D.}~\bibnamefont {Deppe}}, \bibinfo {author} {\bibfnamefont
  {M.}~\bibnamefont {Xiao}}, \bibinfo {author} {\bibfnamefont {W.}~\bibnamefont
  {Ma}}, \bibinfo {author} {\bibfnamefont {G.}~\bibnamefont {Salamo}}, \ and\
  \bibinfo {author} {\bibfnamefont {C.}~\bibnamefont {Shih}},\ }\href@noop {}
  {\bibfield  {journal} {\bibinfo  {journal} {Nat. Phys.}\ }\textbf {\bibinfo
  {volume} {5}},\ \bibinfo {pages} {203} (\bibinfo {year} {2009})}\BibitemShut
  {NoStop}%
\bibitem [{\citenamefont {Gao}\ \emph {et~al.}(2012)\citenamefont {Gao},
  \citenamefont {Fallahi}, \citenamefont {Togan}, \citenamefont
  {Miguel-Sanchez},\ and\ \citenamefont {Imamoglu}}]{Gao12}%
  \BibitemOpen
  \bibfield  {author} {\bibinfo {author} {\bibfnamefont {W.~B.}\ \bibnamefont
  {Gao}}, \bibinfo {author} {\bibfnamefont {P.}~\bibnamefont {Fallahi}},
  \bibinfo {author} {\bibfnamefont {E.}~\bibnamefont {Togan}}, \bibinfo
  {author} {\bibfnamefont {J.}~\bibnamefont {Miguel-Sanchez}}, \ and\ \bibinfo
  {author} {\bibfnamefont {A.}~\bibnamefont {Imamoglu}},\ }\href@noop {}
  {\bibfield  {journal} {\bibinfo  {journal} {Nature}\ }\textbf {\bibinfo
  {volume} {491}},\ \bibinfo {pages} {426} (\bibinfo {year}
  {2012})}\BibitemShut {NoStop}%
\bibitem [{\citenamefont {He}\ \emph {et~al.}(2013)\citenamefont {He},
  \citenamefont {He}, \citenamefont {Wei}, \citenamefont {Wu}, \citenamefont
  {Atature}, \citenamefont {Schneider}, \citenamefont {Hofling}, \citenamefont
  {Kamp}, \citenamefont {Lu},\ and\ \citenamefont {Pan}}]{He13}%
  \BibitemOpen
  \bibfield  {author} {\bibinfo {author} {\bibfnamefont {Y.-M.}\ \bibnamefont
  {He}}, \bibinfo {author} {\bibfnamefont {Y.}~\bibnamefont {He}}, \bibinfo
  {author} {\bibfnamefont {Y.-J.}\ \bibnamefont {Wei}}, \bibinfo {author}
  {\bibfnamefont {D.}~\bibnamefont {Wu}}, \bibinfo {author} {\bibfnamefont
  {M.}~\bibnamefont {Atature}}, \bibinfo {author} {\bibfnamefont
  {C.}~\bibnamefont {Schneider}}, \bibinfo {author} {\bibfnamefont
  {S.}~\bibnamefont {Hofling}}, \bibinfo {author} {\bibfnamefont
  {M.}~\bibnamefont {Kamp}}, \bibinfo {author} {\bibfnamefont {C.-Y.}\
  \bibnamefont {Lu}}, \ and\ \bibinfo {author} {\bibfnamefont {J.-W.}\
  \bibnamefont {Pan}},\ }\href@noop {} {\bibfield  {journal} {\bibinfo
  {journal} {Nat. Nanotechnol.}\ }\textbf {\bibinfo {volume} {8}},\ \bibinfo
  {pages} {213} (\bibinfo {year} {2013})}\BibitemShut {NoStop}%
\bibitem [{\citenamefont {Ellis}\ \emph {et~al.}(2011)\citenamefont {Ellis},
  \citenamefont {Mayer}, \citenamefont {Shambat}, \citenamefont {Sarmiento},
  \citenamefont {Harris}, \citenamefont {Haller},\ and\ \citenamefont
  {Vuckovic}}]{Ellis11}%
  \BibitemOpen
  \bibfield  {author} {\bibinfo {author} {\bibfnamefont {B.}~\bibnamefont
  {Ellis}}, \bibinfo {author} {\bibfnamefont {M.}~\bibnamefont {Mayer}},
  \bibinfo {author} {\bibfnamefont {G.}~\bibnamefont {Shambat}}, \bibinfo
  {author} {\bibfnamefont {T.}~\bibnamefont {Sarmiento}}, \bibinfo {author}
  {\bibfnamefont {J.}~\bibnamefont {Harris}}, \bibinfo {author} {\bibfnamefont
  {E.}~\bibnamefont {Haller}}, \ and\ \bibinfo {author} {\bibfnamefont
  {J.}~\bibnamefont {Vuckovic}},\ }\href@noop {} {\bibfield  {journal}
  {\bibinfo  {journal} {Nat. Phot.}\ }\textbf {\bibinfo {volume} {5}},\
  \bibinfo {pages} {297} (\bibinfo {year} {2011})}\BibitemShut {NoStop}%
\bibitem [{\citenamefont {Ledentsov}\ \emph {et~al.}(1998)\citenamefont
  {Ledentsov}, \citenamefont {Ustinov}, \citenamefont {Shchukin}, \citenamefont
  {Kop'ev}, \citenamefont {Alferov},\ and\ \citenamefont
  {Bimberg}}]{Ledentsov98}%
  \BibitemOpen
  \bibfield  {author} {\bibinfo {author} {\bibfnamefont {N.}~\bibnamefont
  {Ledentsov}}, \bibinfo {author} {\bibfnamefont {V.}~\bibnamefont {Ustinov}},
  \bibinfo {author} {\bibfnamefont {V.}~\bibnamefont {Shchukin}}, \bibinfo
  {author} {\bibfnamefont {P.}~\bibnamefont {Kop'ev}}, \bibinfo {author}
  {\bibfnamefont {Z.}~\bibnamefont {Alferov}}, \ and\ \bibinfo {author}
  {\bibfnamefont {D.}~\bibnamefont {Bimberg}},\ }\href@noop {} {\bibfield
  {journal} {\bibinfo  {journal} {Semiconductors}\ }\textbf {\bibinfo {volume}
  {32}},\ \bibinfo {pages} {343} (\bibinfo {year} {1998})}\BibitemShut
  {NoStop}%
\bibitem [{\citenamefont {Ellis}\ \emph {et~al.}(2007)\citenamefont {Ellis},
  \citenamefont {Fushman}, \citenamefont {Englund}, \citenamefont {Zhang},
  \citenamefont {Yamamoto},\ and\ \citenamefont
  {J.~Vu\v{c}kovi\'{c}}}]{Ellis07}%
  \BibitemOpen
  \bibfield  {author} {\bibinfo {author} {\bibfnamefont {B.}~\bibnamefont
  {Ellis}}, \bibinfo {author} {\bibfnamefont {I.}~\bibnamefont {Fushman}},
  \bibinfo {author} {\bibfnamefont {D.}~\bibnamefont {Englund}}, \bibinfo
  {author} {\bibfnamefont {B.}~\bibnamefont {Zhang}}, \bibinfo {author}
  {\bibfnamefont {Y.}~\bibnamefont {Yamamoto}}, \ and\ \bibinfo {author}
  {\bibfnamefont {J.}~\bibnamefont {J.~Vu\v{c}kovi\'{c}}},\ }\href@noop {}
  {\bibfield  {journal} {\bibinfo  {journal} {Appl. Phys. Lett.}\ }\textbf
  {\bibinfo {volume} {90}},\ \bibinfo {pages} {151102} (\bibinfo {year}
  {2007})}\BibitemShut {NoStop}%
\bibitem [{\citenamefont {Urayama}\ \emph {et~al.}(2001)\citenamefont
  {Urayama}, \citenamefont {Norris}, \citenamefont {Singh},\ and\ \citenamefont
  {Bhattacharya}}]{Urayama01}%
  \BibitemOpen
  \bibfield  {author} {\bibinfo {author} {\bibfnamefont {J.}~\bibnamefont
  {Urayama}}, \bibinfo {author} {\bibfnamefont {T.~B.}\ \bibnamefont {Norris}},
  \bibinfo {author} {\bibfnamefont {J.}~\bibnamefont {Singh}}, \ and\ \bibinfo
  {author} {\bibfnamefont {P.}~\bibnamefont {Bhattacharya}},\ }\href@noop {}
  {\bibfield  {journal} {\bibinfo  {journal} {Phys. Rev. Lett.}\ }\textbf
  {\bibinfo {volume} {86}},\ \bibinfo {pages} {4930} (\bibinfo {year}
  {2001})}\BibitemShut {NoStop}%
\bibitem [{\citenamefont {Benisty}, \citenamefont {Sotomayor-Torr\`es},\ and\
  \citenamefont {Weisbuch}(1991)}]{Benisty91}%
  \BibitemOpen
  \bibfield  {author} {\bibinfo {author} {\bibfnamefont {H.}~\bibnamefont
  {Benisty}}, \bibinfo {author} {\bibfnamefont {C.~M.}\ \bibnamefont
  {Sotomayor-Torr\`es}}, \ and\ \bibinfo {author} {\bibfnamefont
  {C.}~\bibnamefont {Weisbuch}},\ }\href@noop {} {\bibfield  {journal}
  {\bibinfo  {journal} {Phys. Rev. B}\ }\textbf {\bibinfo {volume} {44}},\
  \bibinfo {pages} {10945} (\bibinfo {year} {1991})}\BibitemShut {NoStop}%
\bibitem [{\citenamefont {Gol'tsman}\ \emph {et~al.}(2005)\citenamefont
  {Gol'tsman}, \citenamefont {Korneev}, \citenamefont {Rubtsova}, \citenamefont
  {Milostnaya}, \citenamefont {Chulkova}, \citenamefont {Minaeva},
  \citenamefont {Smirnov}, \citenamefont {Voronov}, \citenamefont {Slysz},
  \citenamefont {Pearlman} \emph {et~al.}}]{Goltsman05}%
  \BibitemOpen
  \bibfield  {author} {\bibinfo {author} {\bibfnamefont {G.~N.}\ \bibnamefont
  {Gol'tsman}}, \bibinfo {author} {\bibfnamefont {A.}~\bibnamefont {Korneev}},
  \bibinfo {author} {\bibfnamefont {I.}~\bibnamefont {Rubtsova}}, \bibinfo
  {author} {\bibfnamefont {I.}~\bibnamefont {Milostnaya}}, \bibinfo {author}
  {\bibfnamefont {G.}~\bibnamefont {Chulkova}}, \bibinfo {author}
  {\bibfnamefont {O.}~\bibnamefont {Minaeva}}, \bibinfo {author} {\bibfnamefont
  {K.}~\bibnamefont {Smirnov}}, \bibinfo {author} {\bibfnamefont
  {B.}~\bibnamefont {Voronov}}, \bibinfo {author} {\bibfnamefont
  {W.}~\bibnamefont {Slysz}}, \bibinfo {author} {\bibfnamefont
  {A.}~\bibnamefont {Pearlman}},  \emph {et~al.},\ }\href@noop {} {\bibfield
  {journal} {\bibinfo  {journal} {Phys. Stat. Sol. C}\ }\textbf {\bibinfo
  {volume} {2}},\ \bibinfo {pages} {1480} (\bibinfo {year} {2005})}\BibitemShut
  {NoStop}%
\bibitem [{\citenamefont {Najafi}\ \emph {et~al.}(2012)\citenamefont {Najafi},
  \citenamefont {Marsili}, \citenamefont {Dauler}, \citenamefont {Molnar},\
  and\ \citenamefont {Berggren}}]{Najafi12}%
  \BibitemOpen
  \bibfield  {author} {\bibinfo {author} {\bibfnamefont {F.}~\bibnamefont
  {Najafi}}, \bibinfo {author} {\bibfnamefont {F.}~\bibnamefont {Marsili}},
  \bibinfo {author} {\bibfnamefont {E.}~\bibnamefont {Dauler}}, \bibinfo
  {author} {\bibfnamefont {R.}~\bibnamefont {Molnar}}, \ and\ \bibinfo {author}
  {\bibfnamefont {K.}~\bibnamefont {Berggren}},\ }\href@noop {} {\bibfield
  {journal} {\bibinfo  {journal} {Appl. Phys. Lett.}\ }\textbf {\bibinfo
  {volume} {100}},\ \bibinfo {pages} {152602} (\bibinfo {year}
  {2012})}\BibitemShut {NoStop}%
\bibitem [{\citenamefont {Pernice}\ \emph {et~al.}(2012)\citenamefont
  {Pernice}, \citenamefont {Schuck}, \citenamefont {Minaeva}, \citenamefont
  {Li}, \citenamefont {Goltsman}, \citenamefont {Sergienko},\ and\
  \citenamefont {Tang}}]{Pernice12}%
  \BibitemOpen
  \bibfield  {author} {\bibinfo {author} {\bibfnamefont {W.}~\bibnamefont
  {Pernice}}, \bibinfo {author} {\bibfnamefont {C.}~\bibnamefont {Schuck}},
  \bibinfo {author} {\bibfnamefont {O.}~\bibnamefont {Minaeva}}, \bibinfo
  {author} {\bibfnamefont {M.}~\bibnamefont {Li}}, \bibinfo {author}
  {\bibfnamefont {G.}~\bibnamefont {Goltsman}}, \bibinfo {author}
  {\bibfnamefont {A.}~\bibnamefont {Sergienko}}, \ and\ \bibinfo {author}
  {\bibfnamefont {H.}~\bibnamefont {Tang}},\ }\href@noop {} {\bibfield
  {journal} {\bibinfo  {journal} {Nat. Comm.}\ }\textbf {\bibinfo {volume}
  {3}},\ \bibinfo {eid} {1325} (\bibinfo {year} {2012})}\BibitemShut {NoStop}%
\bibitem [{\citenamefont {Sprengers}\ \emph {et~al.}(2011)\citenamefont
  {Sprengers}, \citenamefont {Gaggero}, \citenamefont {Sahin}, \citenamefont
  {Jahanmirinejad}, \citenamefont {Frucci}, \citenamefont {Mattioli},
  \citenamefont {Leoni}, \citenamefont {Beetz}, \citenamefont {Lermer},
  \citenamefont {Kamp}, \citenamefont {Hofling}, \citenamefont {Sanjines},\
  and\ \citenamefont {Fiore}}]{Sprengers11}%
  \BibitemOpen
  \bibfield  {author} {\bibinfo {author} {\bibfnamefont {J.~P.}\ \bibnamefont
  {Sprengers}}, \bibinfo {author} {\bibfnamefont {A.}~\bibnamefont {Gaggero}},
  \bibinfo {author} {\bibfnamefont {D.}~\bibnamefont {Sahin}}, \bibinfo
  {author} {\bibfnamefont {S.}~\bibnamefont {Jahanmirinejad}}, \bibinfo
  {author} {\bibfnamefont {G.}~\bibnamefont {Frucci}}, \bibinfo {author}
  {\bibfnamefont {F.}~\bibnamefont {Mattioli}}, \bibinfo {author}
  {\bibfnamefont {R.}~\bibnamefont {Leoni}}, \bibinfo {author} {\bibfnamefont
  {J.}~\bibnamefont {Beetz}}, \bibinfo {author} {\bibfnamefont
  {M.}~\bibnamefont {Lermer}}, \bibinfo {author} {\bibfnamefont
  {M.}~\bibnamefont {Kamp}}, \bibinfo {author} {\bibfnamefont {S.}~\bibnamefont
  {Hofling}}, \bibinfo {author} {\bibfnamefont {R.}~\bibnamefont {Sanjines}}, \
  and\ \bibinfo {author} {\bibfnamefont {A.}~\bibnamefont {Fiore}},\ }\href
  {\doibase 10.1063/1.3657518} {\bibfield  {journal} {\bibinfo  {journal}
  {Appl. Phys. Lett.}\ }\textbf {\bibinfo {volume} {99}},\ \bibinfo {pages}
  {181110} (\bibinfo {year} {2011})}\BibitemShut {NoStop}%
\bibitem [{\citenamefont {Sahin}\ \emph {et~al.}(2013)\citenamefont {Sahin},
  \citenamefont {Gaggero}, \citenamefont {Hoang}, \citenamefont {Frucci},
  \citenamefont {Mattioli}, \citenamefont {Leoni}, \citenamefont {Beetz},
  \citenamefont {Lermer}, \citenamefont {Kamp}, \citenamefont {H\"{o}fling}
  \emph {et~al.}}]{Sahin13}%
  \BibitemOpen
  \bibfield  {author} {\bibinfo {author} {\bibfnamefont {D.}~\bibnamefont
  {Sahin}}, \bibinfo {author} {\bibfnamefont {A.}~\bibnamefont {Gaggero}},
  \bibinfo {author} {\bibfnamefont {T.~B.}\ \bibnamefont {Hoang}}, \bibinfo
  {author} {\bibfnamefont {G.}~\bibnamefont {Frucci}}, \bibinfo {author}
  {\bibfnamefont {F.}~\bibnamefont {Mattioli}}, \bibinfo {author}
  {\bibfnamefont {R.}~\bibnamefont {Leoni}}, \bibinfo {author} {\bibfnamefont
  {J.}~\bibnamefont {Beetz}}, \bibinfo {author} {\bibfnamefont
  {M.}~\bibnamefont {Lermer}}, \bibinfo {author} {\bibfnamefont
  {M.}~\bibnamefont {Kamp}}, \bibinfo {author} {\bibfnamefont {S.}~\bibnamefont
  {H\"{o}fling}},  \emph {et~al.},\ }\href {\doibase 10.1364/OE.21.011162}
  {\bibfield  {journal} {\bibinfo  {journal} {Opt. Express}\ }\textbf {\bibinfo
  {volume} {21}},\ \bibinfo {pages} {11162} (\bibinfo {year}
  {2013})}\BibitemShut {NoStop}%
\bibitem [{\citenamefont {Marsili}\ \emph {et~al.}(2008)\citenamefont
  {Marsili}, \citenamefont {Bitauld}, \citenamefont {Fiore}, \citenamefont
  {Gaggero}, \citenamefont {Mattioli}, \citenamefont {Leoni}, \citenamefont
  {Benkahoul},\ and\ \citenamefont {L\'{e}vy}}]{Marsili08}%
  \BibitemOpen
  \bibfield  {author} {\bibinfo {author} {\bibfnamefont {F.}~\bibnamefont
  {Marsili}}, \bibinfo {author} {\bibfnamefont {D.}~\bibnamefont {Bitauld}},
  \bibinfo {author} {\bibfnamefont {A.}~\bibnamefont {Fiore}}, \bibinfo
  {author} {\bibfnamefont {A.}~\bibnamefont {Gaggero}}, \bibinfo {author}
  {\bibfnamefont {F.}~\bibnamefont {Mattioli}}, \bibinfo {author}
  {\bibfnamefont {R.}~\bibnamefont {Leoni}}, \bibinfo {author} {\bibfnamefont
  {M.}~\bibnamefont {Benkahoul}}, \ and\ \bibinfo {author} {\bibfnamefont
  {F.}~\bibnamefont {L\'{e}vy}},\ }\href@noop {} {\bibfield  {journal}
  {\bibinfo  {journal} {Opt. Express}\ }\textbf {\bibinfo {volume} {16}},\
  \bibinfo {pages} {3191} (\bibinfo {year} {2008})}\BibitemShut {NoStop}%
\bibitem [{\citenamefont {Kerman}\ \emph {et~al.}(2007)\citenamefont {Kerman},
  \citenamefont {Dauler}, \citenamefont {Yang}, \citenamefont {Rosfjord},
  \citenamefont {Anant}, \citenamefont {Berggren}, \citenamefont
  {Gol'tsman},\ and\ \citenamefont {Voronov}}]{Kerman07}%
  \BibitemOpen
  \bibfield  {author} {\bibinfo {author} {\bibfnamefont {A.~J.}\ \bibnamefont
  {Kerman}}, \bibinfo {author} {\bibfnamefont {E.~A.}\ \bibnamefont {Dauler}},
  \bibinfo {author} {\bibfnamefont {J.~K.~W.}\ \bibnamefont {Yang}}, \bibinfo
  {author} {\bibfnamefont {K.~M.}\ \bibnamefont {Rosfjord}}, \bibinfo {author}
  {\bibfnamefont {V.}~\bibnamefont {Anant}}, \bibinfo {author} {\bibfnamefont
  {K.~K.}\ \bibnamefont {Berggren}}, \bibinfo {author} {\bibfnamefont {G.~N.}\
  \bibnamefont {Gol'tsman}}, \ and\ \bibinfo {author} {\bibfnamefont {B.~M.}\
  \bibnamefont {Voronov}},\ }\href {\doibase
  http://dx.doi.org/10.1063/1.2696926} {\bibfield  {journal} {\bibinfo
  {journal} {Appl. Phys. Lett.}\ }\textbf {\bibinfo {volume} {90}},\ \bibinfo
  {eid} {101110} (\bibinfo {year} {2007})}\BibitemShut {NoStop}%
\bibitem [{\citenamefont {Reithmaier}\ \emph
  {et~al.}(2013{\natexlab{a}})\citenamefont {Reithmaier}, \citenamefont {Senf},
  \citenamefont {Lichtmannecker}, \citenamefont {Reichert}, \citenamefont
  {Flassig}, \citenamefont {Voss}, \citenamefont {Gross},\ and\ \citenamefont
  {Finley}}]{Reithmaier13}%
  \BibitemOpen
  \bibfield  {author} {\bibinfo {author} {\bibfnamefont {G.}~\bibnamefont
  {Reithmaier}}, \bibinfo {author} {\bibfnamefont {J.}~\bibnamefont {Senf}},
  \bibinfo {author} {\bibfnamefont {S.}~\bibnamefont {Lichtmannecker}},
  \bibinfo {author} {\bibfnamefont {T.}~\bibnamefont {Reichert}}, \bibinfo
  {author} {\bibfnamefont {F.}~\bibnamefont {Flassig}}, \bibinfo {author}
  {\bibfnamefont {A.}~\bibnamefont {Voss}}, \bibinfo {author} {\bibfnamefont
  {R.}~\bibnamefont {Gross}}, \ and\ \bibinfo {author} {\bibfnamefont {J.~J.}\
  \bibnamefont {Finley}},\ }\href@noop {} {\bibfield  {journal} {\bibinfo
  {journal} {J. Appl. Phys.}\ }\textbf {\bibinfo {volume} {113}},\ \bibinfo
  {eid} {143507} (\bibinfo {year} {2013}{\natexlab{a}})}\BibitemShut {NoStop}%
\bibitem [{\citenamefont {Reithmaier}\ \emph
  {et~al.}(2013{\natexlab{b}})\citenamefont {Reithmaier}, \citenamefont
  {Lichtmannecker}, \citenamefont {Reichert}, \citenamefont {Hasch},
  \citenamefont {M\"uller}, \citenamefont {Bichler}, \citenamefont {Gross},\
  and\ \citenamefont {Finley}}]{Reithmaier13SciRep}%
  \BibitemOpen
  \bibfield  {author} {\bibinfo {author} {\bibfnamefont {G.}~\bibnamefont
  {Reithmaier}}, \bibinfo {author} {\bibfnamefont {S.}~\bibnamefont
  {Lichtmannecker}}, \bibinfo {author} {\bibfnamefont {T.}~\bibnamefont
  {Reichert}}, \bibinfo {author} {\bibfnamefont {P.}~\bibnamefont {Hasch}},
  \bibinfo {author} {\bibfnamefont {K.}~\bibnamefont {M\"uller}}, \bibinfo
  {author} {\bibfnamefont {M.}~\bibnamefont {Bichler}}, \bibinfo {author}
  {\bibfnamefont {R.}~\bibnamefont {Gross}}, \ and\ \bibinfo {author}
  {\bibfnamefont {J.~J.}\ \bibnamefont {Finley}},\ }\href@noop {} {\bibfield
  {journal} {\bibinfo  {journal} {Sci. Rep.}\ }\textbf {\bibinfo {volume}
  {3}},\ \bibinfo {eid} {1901} (\bibinfo {year}
  {2013}{\natexlab{b}})}\BibitemShut {NoStop}%
\bibitem [{\citenamefont {Finley}\ \emph {et~al.}(2001)\citenamefont {Finley},
  \citenamefont {Ashmore}, \citenamefont {Lemaitre}, \citenamefont {Mowbray},
  \citenamefont {Skolnick}, \citenamefont {Itskevich}, \citenamefont {Maksym},
  \citenamefont {Hopkinson},\ and\ \citenamefont {Krauss}}]{Finley01}%
  \BibitemOpen
  \bibfield  {author} {\bibinfo {author} {\bibfnamefont {J.}~\bibnamefont
  {Finley}}, \bibinfo {author} {\bibfnamefont {A.}~\bibnamefont {Ashmore}},
  \bibinfo {author} {\bibfnamefont {A.}~\bibnamefont {Lemaitre}}, \bibinfo
  {author} {\bibfnamefont {D.}~\bibnamefont {Mowbray}}, \bibinfo {author}
  {\bibfnamefont {M.}~\bibnamefont {Skolnick}}, \bibinfo {author}
  {\bibfnamefont {I.}~\bibnamefont {Itskevich}}, \bibinfo {author}
  {\bibfnamefont {P.}~\bibnamefont {Maksym}}, \bibinfo {author} {\bibfnamefont
  {M.}~\bibnamefont {Hopkinson}}, \ and\ \bibinfo {author} {\bibfnamefont
  {T.}~\bibnamefont {Krauss}},\ }\href {\doibase 10.1103/PhysRevB.63.073307}
  {\bibfield  {journal} {\bibinfo  {journal} {Phys. Rev. B}\ }\textbf {\bibinfo
  {volume} {63}},\ \bibinfo {pages} {073307} (\bibinfo {year}
  {2001})}\BibitemShut {NoStop}%
\bibitem [{\citenamefont {Vasanelli}, \citenamefont {Ferreira},\ and\
  \citenamefont {Bastard}(2002)}]{Vasanelli02}%
  \BibitemOpen
  \bibfield  {author} {\bibinfo {author} {\bibfnamefont {A.}~\bibnamefont
  {Vasanelli}}, \bibinfo {author} {\bibfnamefont {R.}~\bibnamefont {Ferreira}},
  \ and\ \bibinfo {author} {\bibfnamefont {G.}~\bibnamefont {Bastard}},\
  }\href@noop {} {\bibfield  {journal} {\bibinfo  {journal} {Phys. Rev. Lett.}\
  }\textbf {\bibinfo {volume} {89}},\ \bibinfo {pages} {216804} (\bibinfo
  {year} {2002})}\BibitemShut {NoStop}%
\bibitem [{Note1()}]{Note1}%
  \BibitemOpen
  \bibinfo {note} {The overall temporal resolution of the on-chip experiment is
  limited to $< 70$ ps, reached by deconvolution of the signal being recorded
  using a Picoquant Timeharp 200 that exhibits a timing uncertainty of $< 150$
  ps.}\BibitemShut {Stop}%
\bibitem [{\citenamefont {Viasnoff-Schwoob}\ \emph {et~al.}(2005)\citenamefont
  {Viasnoff-Schwoob}, \citenamefont {Weisbuch}, \citenamefont {Benisty},
  \citenamefont {Olivier}, \citenamefont {Varoutsis}, \citenamefont
  {Robert-Philip}, \citenamefont {Houdr\'e},\ and\ \citenamefont
  {Smith}}]{Schwoob05}%
  \BibitemOpen
  \bibfield  {author} {\bibinfo {author} {\bibfnamefont {E.}~\bibnamefont
  {Viasnoff-Schwoob}}, \bibinfo {author} {\bibfnamefont {C.}~\bibnamefont
  {Weisbuch}}, \bibinfo {author} {\bibfnamefont {H.}~\bibnamefont {Benisty}},
  \bibinfo {author} {\bibfnamefont {S.}~\bibnamefont {Olivier}}, \bibinfo
  {author} {\bibfnamefont {S.}~\bibnamefont {Varoutsis}}, \bibinfo {author}
  {\bibfnamefont {I.}~\bibnamefont {Robert-Philip}}, \bibinfo {author}
  {\bibfnamefont {R.}~\bibnamefont {Houdr\'e}}, \ and\ \bibinfo {author}
  {\bibfnamefont {C.~J.~M.}\ \bibnamefont {Smith}},\ }\href {\doibase
  10.1103/PhysRevLett.95.183901} {\bibfield  {journal} {\bibinfo  {journal}
  {Phys. Rev. Lett.}\ }\textbf {\bibinfo {volume} {95}},\ \bibinfo {pages}
  {183901} (\bibinfo {year} {2005})}\BibitemShut {NoStop}%
\bibitem [{\citenamefont {Rao}\ and\ \citenamefont {Hughes}(2007)}]{Rao07}%
  \BibitemOpen
  \bibfield  {author} {\bibinfo {author} {\bibfnamefont {V.~S. C.~M.}\
  \bibnamefont {Rao}}\ and\ \bibinfo {author} {\bibfnamefont {S.}~\bibnamefont
  {Hughes}},\ }\href {\doibase 10.1103/PhysRevLett.99.193901} {\bibfield
  {journal} {\bibinfo  {journal} {Phys. Rev. Lett.}\ }\textbf {\bibinfo
  {volume} {99}},\ \bibinfo {pages} {193901} (\bibinfo {year}
  {2007})}\BibitemShut {NoStop}%
\bibitem [{\citenamefont {Santori}\ \emph {et~al.}(2002)\citenamefont
  {Santori}, \citenamefont {Solomon}, \citenamefont {Pelton},\ and\
  \citenamefont {Yamamoto}}]{Santori02}%
  \BibitemOpen
  \bibfield  {author} {\bibinfo {author} {\bibfnamefont {C.}~\bibnamefont
  {Santori}}, \bibinfo {author} {\bibfnamefont {G.~S.}\ \bibnamefont
  {Solomon}}, \bibinfo {author} {\bibfnamefont {M.}~\bibnamefont {Pelton}}, \
  and\ \bibinfo {author} {\bibfnamefont {Y.}~\bibnamefont {Yamamoto}},\ }\href
  {\doibase 10.1103/PhysRevB.65.073310} {\bibfield  {journal} {\bibinfo
  {journal} {Phys. Rev. B}\ }\textbf {\bibinfo {volume} {65}},\ \bibinfo
  {pages} {073310} (\bibinfo {year} {2002})}\BibitemShut {NoStop}%
\bibitem [{\citenamefont {Raymond}\ \emph {et~al.}(1996)\citenamefont
  {Raymond}, \citenamefont {Fafard}, \citenamefont {Poole}, \citenamefont
  {Wojs}, \citenamefont {Hawrylak}, \citenamefont {Charbonneau}, \citenamefont
  {Leonard}, \citenamefont {Leon}, \citenamefont {Petroff},\ and\ \citenamefont
  {Merz}}]{Raymond96}%
  \BibitemOpen
  \bibfield  {author} {\bibinfo {author} {\bibfnamefont {S.}~\bibnamefont
  {Raymond}}, \bibinfo {author} {\bibfnamefont {S.}~\bibnamefont {Fafard}},
  \bibinfo {author} {\bibfnamefont {P.~J.}\ \bibnamefont {Poole}}, \bibinfo
  {author} {\bibfnamefont {A.}~\bibnamefont {Wojs}}, \bibinfo {author}
  {\bibfnamefont {P.}~\bibnamefont {Hawrylak}}, \bibinfo {author}
  {\bibfnamefont {S.}~\bibnamefont {Charbonneau}}, \bibinfo {author}
  {\bibfnamefont {D.}~\bibnamefont {Leonard}}, \bibinfo {author} {\bibfnamefont
  {R.}~\bibnamefont {Leon}}, \bibinfo {author} {\bibfnamefont {P.~M.}\
  \bibnamefont {Petroff}}, \ and\ \bibinfo {author} {\bibfnamefont {J.~L.}\
  \bibnamefont {Merz}},\ }\href {\doibase 10.1103/PhysRevB.54.11548} {\bibfield
   {journal} {\bibinfo  {journal} {Phys. Rev. B}\ }\textbf {\bibinfo {volume}
  {54}},\ \bibinfo {pages} {11548} (\bibinfo {year} {1996})}\BibitemShut
  {NoStop}%
\bibitem [{\citenamefont {M'gha\"{\i}eth}\ \emph {et~al.}(1999)\citenamefont
  {M'gha\"{\i}eth}, \citenamefont {Ma\^aref}, \citenamefont {Mihalcescu},\ and\
  \citenamefont {Vial}}]{Mghaieth99}%
  \BibitemOpen
  \bibfield  {author} {\bibinfo {author} {\bibfnamefont {R.}~\bibnamefont
  {M'gha\"{\i}eth}}, \bibinfo {author} {\bibfnamefont {H.}~\bibnamefont
  {Ma\^aref}}, \bibinfo {author} {\bibfnamefont {I.}~\bibnamefont
  {Mihalcescu}}, \ and\ \bibinfo {author} {\bibfnamefont {J.~C.}\ \bibnamefont
  {Vial}},\ }\href@noop {} {\bibfield  {journal} {\bibinfo  {journal} {Phys.
  Rev. B}\ }\textbf {\bibinfo {volume} {60}},\ \bibinfo {pages} {4450}
  (\bibinfo {year} {1999})}\BibitemShut {NoStop}%
\bibitem [{\citenamefont {Steinhoff}\ \emph {et~al.}(2013)\citenamefont
  {Steinhoff}, \citenamefont {Kurtze}, \citenamefont {Gartner}, \citenamefont
  {Florian}, \citenamefont {Reuter}, \citenamefont {Wieck}, \citenamefont
  {Bayer},\ and\ \citenamefont {Jahnke}}]{Steinhoff13}%
  \BibitemOpen
  \bibfield  {author} {\bibinfo {author} {\bibfnamefont {A.}~\bibnamefont
  {Steinhoff}}, \bibinfo {author} {\bibfnamefont {H.}~\bibnamefont {Kurtze}},
  \bibinfo {author} {\bibfnamefont {P.}~\bibnamefont {Gartner}}, \bibinfo
  {author} {\bibfnamefont {M.}~\bibnamefont {Florian}}, \bibinfo {author}
  {\bibfnamefont {D.}~\bibnamefont {Reuter}}, \bibinfo {author} {\bibfnamefont
  {A.~D.}\ \bibnamefont {Wieck}}, \bibinfo {author} {\bibfnamefont
  {M.}~\bibnamefont {Bayer}}, \ and\ \bibinfo {author} {\bibfnamefont
  {F.}~\bibnamefont {Jahnke}},\ }\href@noop {} {\bibfield  {journal} {\bibinfo
  {journal} {Phys. Rev. B}\ }\textbf {\bibinfo {volume} {88}},\ \bibinfo
  {pages} {205309} (\bibinfo {year} {2013})}\BibitemShut {NoStop}%
\bibitem [{\citenamefont {Bacher}\ \emph {et~al.}(1999)\citenamefont {Bacher},
  \citenamefont {Weigand}, \citenamefont {Seufert}, \citenamefont
  {Kulakovskii}, \citenamefont {Gippius}, \citenamefont {Forchel},
  \citenamefont {Leonardi},\ and\ \citenamefont {Hommel}}]{Bacher99}%
  \BibitemOpen
  \bibfield  {author} {\bibinfo {author} {\bibfnamefont {G.}~\bibnamefont
  {Bacher}}, \bibinfo {author} {\bibfnamefont {R.}~\bibnamefont {Weigand}},
  \bibinfo {author} {\bibfnamefont {J.}~\bibnamefont {Seufert}}, \bibinfo
  {author} {\bibfnamefont {V.~D.}\ \bibnamefont {Kulakovskii}}, \bibinfo
  {author} {\bibfnamefont {N.~A.}\ \bibnamefont {Gippius}}, \bibinfo {author}
  {\bibfnamefont {A.}~\bibnamefont {Forchel}}, \bibinfo {author} {\bibfnamefont
  {K.}~\bibnamefont {Leonardi}}, \ and\ \bibinfo {author} {\bibfnamefont
  {D.}~\bibnamefont {Hommel}},\ }\href@noop {} {\bibfield  {journal} {\bibinfo
  {journal} {Phys. Rev. Lett.}\ }\textbf {\bibinfo {volume} {83}},\ \bibinfo
  {pages} {4417} (\bibinfo {year} {1999})}\BibitemShut {NoStop}%
\bibitem [{\citenamefont {Ikeda}\ \emph {et~al.}(2004)\citenamefont {Ikeda},
  \citenamefont {Sekiguchi}, \citenamefont {Minami}, \citenamefont {Yoshino},
  \citenamefont {Mitsumori}, \citenamefont {Amanai}, \citenamefont {Nagao},\
  and\ \citenamefont {Sakaki}}]{Ikeda04}%
  \BibitemOpen
  \bibfield  {author} {\bibinfo {author} {\bibfnamefont {K.}~\bibnamefont
  {Ikeda}}, \bibinfo {author} {\bibfnamefont {H.}~\bibnamefont {Sekiguchi}},
  \bibinfo {author} {\bibfnamefont {F.}~\bibnamefont {Minami}}, \bibinfo
  {author} {\bibfnamefont {J.}~\bibnamefont {Yoshino}}, \bibinfo {author}
  {\bibfnamefont {Y.}~\bibnamefont {Mitsumori}}, \bibinfo {author}
  {\bibfnamefont {H.}~\bibnamefont {Amanai}}, \bibinfo {author} {\bibfnamefont
  {S.}~\bibnamefont {Nagao}}, \ and\ \bibinfo {author} {\bibfnamefont
  {S.}~\bibnamefont {Sakaki}},\ }\href@noop {} {\bibfield  {journal} {\bibinfo
  {journal} {J. Lumin.}\ }\textbf {\bibinfo {volume} {108}},\ \bibinfo {pages}
  {273 } (\bibinfo {year} {2004})}\BibitemShut {NoStop}%
\bibitem [{\citenamefont {Heitz}\ \emph {et~al.}(2001)\citenamefont {Heitz},
  \citenamefont {Born}, \citenamefont {Guffarth}, \citenamefont {Stier},
  \citenamefont {Schliwa}, \citenamefont {Hoffmann},\ and\ \citenamefont
  {Bimberg}}]{Heitz01}%
  \BibitemOpen
  \bibfield  {author} {\bibinfo {author} {\bibfnamefont {R.}~\bibnamefont
  {Heitz}}, \bibinfo {author} {\bibfnamefont {H.}~\bibnamefont {Born}},
  \bibinfo {author} {\bibfnamefont {F.}~\bibnamefont {Guffarth}}, \bibinfo
  {author} {\bibfnamefont {O.}~\bibnamefont {Stier}}, \bibinfo {author}
  {\bibfnamefont {A.}~\bibnamefont {Schliwa}}, \bibinfo {author} {\bibfnamefont
  {A.}~\bibnamefont {Hoffmann}}, \ and\ \bibinfo {author} {\bibfnamefont
  {D.}~\bibnamefont {Bimberg}},\ }\href@noop {} {\bibfield  {journal} {\bibinfo
   {journal} {Phys. Rev. B}\ }\textbf {\bibinfo {volume} {64}},\ \bibinfo
  {pages} {241305} (\bibinfo {year} {2001})}\BibitemShut {NoStop}%
\bibitem [{\citenamefont {Kitaygorsky}\ \emph {et~al.}(2007)\citenamefont
  {Kitaygorsky}, \citenamefont {Komissarov}, \citenamefont {Jukna},
  \citenamefont {Pan}, \citenamefont {Minaeva}, \citenamefont {Kaurova},
  \citenamefont {Divochiy}, \citenamefont {Korneev}, \citenamefont {Tarkhov},
  \citenamefont {Voronov} \emph {et~al.}}]{Kitaygorsky07}%
  \BibitemOpen
  \bibfield  {author} {\bibinfo {author} {\bibfnamefont {J.}~\bibnamefont
  {Kitaygorsky}}, \bibinfo {author} {\bibfnamefont {I.}~\bibnamefont
  {Komissarov}}, \bibinfo {author} {\bibfnamefont {A.}~\bibnamefont {Jukna}},
  \bibinfo {author} {\bibfnamefont {D.}~\bibnamefont {Pan}}, \bibinfo {author}
  {\bibfnamefont {O.}~\bibnamefont {Minaeva}}, \bibinfo {author} {\bibfnamefont
  {N.}~\bibnamefont {Kaurova}}, \bibinfo {author} {\bibfnamefont
  {A.}~\bibnamefont {Divochiy}}, \bibinfo {author} {\bibfnamefont
  {A.}~\bibnamefont {Korneev}}, \bibinfo {author} {\bibfnamefont
  {M.}~\bibnamefont {Tarkhov}}, \bibinfo {author} {\bibfnamefont
  {B.}~\bibnamefont {Voronov}},  \emph {et~al.},\ }\href@noop {} {\bibfield
  {journal} {\bibinfo  {journal} {IEEE Trans. Appl. Supercond.}\ }\textbf
  {\bibinfo {volume} {17}},\ \bibinfo {pages} {275} (\bibinfo {year}
  {2007})}\BibitemShut {NoStop}%
\end{thebibliography}

%

\end{document}